\documentclass[a4wide,amsmath,epsfig,amssymb]{article}
\usepackage{amssymb,amsmath,epsfig}
\usepackage{graphics}
\usepackage{graphicx}
\usepackage{setspace} 
\usepackage{color} 
\newcommand{\matrixx}{\cal{X}}
\newcommand{\vecx}{\vec{x}}
\newcommand{\vecX}{\vec{X}}
\newcommand{\mydefine}{\stackrel{def}{=}}
\newcommand{\mycolor}{black} 
\newcommand{\melf}{\color{\mycolor}{\phi_{f}}}
\newcommand{\linearf}{\color{\mycolor}{l_{f}}}

\def\reg{{\rm\ooalign{\hfil
     \raise.07ex\hbox{\scriptsize R}\hfil\crcr\mathhexbox20D}}}

\title{Choice of Mel Filter Bank in Computing MFCC 
of a Resampled Speech}

\makeatletter
\def\name#1{\gdef\@name{#1\\}}
\makeatother
\author{Laxmi Narayana M, Sunil Kumar Kopparapu \\
TCS Innovation Lab - Mumbai, Tata Consultancy Services, \\
Yantra Park, Thane (West), Maharastra, India. \\
Laxmi.Narayana@tcs.com, SunilKumar.Kopparapu@tcs.com}

\begin{document}
\maketitle
%%%%%%%%%%%%%%%%%%%%%%%%%%%%%%%%%%%%%%%%%%%%%%%%%%%%%%%%%%%%%%%%%%%%%%%%%%%%%%%%%%%%%%%%%%%%%
\begin{abstract}
%%%%%%%%%%%%%%%%%%%%%%%%%%%%%%%%%%%%%%%%%%%%%%%%%%%%%%%%%%%%%%%%%%%%%%%%%%%%%%%%%%%%%%%%%%%%

Mel Frequency Cepstral Coefficients (MFCCs) 
are the most popularly used speech features in
most speech and speaker recognition applications. 
In this paper, we study the effect of resampling a speech signal on these
speech features. 
We first derive a relationship between the MFCC parameters of 
the resampled speech and the MFCC parameters of the original speech. 
We propose six methods of calculating the MFCC parameters of 
downsampled speech by transforming the  Mel filter bank used to compute MFCC
of the original speech. 
We then experimentally compute the MFCC parameters of the down sampled speech 
using the proposed methods and compute the Pearson coefficient 
between the MFCC parameters of the 
downsampled speech and that of the original speech to identify the most
effective choice of Mel-filter band that enables the computed MFCC of the
resampled speech to be as close as possible to the original speech sample MFCC.
%We determine the best among the six methods which gives the best correlation. 

%We substantiate the derived results experimentally, primarily focusing on
%downsampled speech. % I commented this because our approach is different now
%\comment{We calculate the error between the MFCC parameters of the original speech and that of 
%the down sampled speech and the error between the MFCC parameters of the original speech and that
%are obtained from the derived relationship. We show that the latter is lesser
%than the former. }
\end{abstract}

\noindent{\bf Index Terms}: MFCC, Time scale modification, time compression, time expansion.

%%%%%%%%%%%%%%%%%%%%%%%%%%%%%%%%%%%%%%%%%%%%%%%%%%%%%%%%%%%%%%%%%%%%%%%%%%%%%%%%%%%%%%%%%%%%
\section{Introduction}
%%%%%%%%%%%%%%%%%%%%%%%%%%%%%%%%%%%%%%%%%%%%%%%%%%%%%%%%%%%%%%%%%%%%%%%%%%%%%%%%%%%%%%%%%%%%
Time scale modification (TSM) is a class of algorithms that change the playback time of speech/audio signals. By increasing or decreasing the apparent rate of articulation, TSM on one hand, 
is useful to make degraded speech more intelligible and on the other hand, reduces the time needed for a 
listener to listen to a message. Reducing the playback time of speech or {\em
time compression of speech signal} has a variety of applications 
that include teaching aids to the disabled and in human-computer interfaces. 
Time-compressed speech is also referred to as accelerated, compressed, time-scale modified, sped-up, 
rate-converted, or time-altered speech. 
%It is desirable to be able to play back audio and speech material at high speed \cite{wong90} \comment{Why?}. 
Studies have indicated that listening to teaching materials twice that have been speeded up by a 
factor of two is more effective than listening to them once at normal speed \cite{barry92}. 
Time compression techniques have also been used in speech recognition systems to time 
normalize input utterances to a standard length. One potential application is that TSM is often used to 
adjust Radio commercials and the audio of television advertisements to fit
exactly into the $30$ or $60$ seconds. 
Time compression of speech also saves 
storage space and transmission bandwidth for speech messages. 
Time compressed speech has been used to speed up message presentation in voice
mail systems \cite{hejn90}. 
%\getinfo{What about in advertisements?}

In general, time scale modification of a speech signal 
is associated with a parameter called time scale modification (TSM) 
factor or scaling factor. 
In this paper we denote the TSM factor by $\alpha$. 
There are a variety of techniques for time scaling of speech out of which, 
resampling is one of the simplest techniques. 
Resampling of digital signals is basically a process of decimation (for time compression, $\alpha>1$) or 
interpolation (for time expansion, $\alpha<1$) or a combination of both. 
Usually, for decimation, the input signal is 
%low-pass filtered \comment{We are not doing this; may be we
%should mention this somewhere} to prevent aliasing and then 
sub-sampled. 
For interpolation, {\em zeros} are inserted between samples of the original 
input signal.
%before the zero-padded signal is low-pass filtered to remove the mirror bands introduced by the zero padding
For a discrete time signal $x[n]$ the restriction on the TSM factor $\alpha$ to
obtain $x[\alpha n]$ is
that $\alpha$  be a rational number. 
For any $\alpha = \frac{p}{q}$ where $p$ 
and $q$ are integers the signal $x[\alpha n]$ is constructed by 
first interpolating $x[n]$ by a factor of $p$, say $x^p = x[n \uparrow p]$ and then
decimating $x[n]$ by a factor of $q$, namely, $x^q = x[n \downarrow q]$. %\review{}.  
It should be noted that, usually interpolation is carried out before decimation
to eliminate information loss in the pre-filtering of decimation. 

Most often, cepstral features are the 
speech features of choice for many speaker and speech recognition 
systems. For example, the Mel-frequency cepstral coefficient (MFCC) 
\cite{merm80} representation of speech is probably the most commonly 
used representation in speaker recognition and and speech recognition applications
\cite{reyn95, %reyn295, 
rash04, sedd04}. In general, cepstral features are 
more compact, discriminable, and most importantly, nearly 
decorrelated such that they allow the diagonal covariance to be used by 
the hidden Markov models (HMMs) effectively. Therefore, they can usually 
provide higher baseline performance over filter bank features 
\cite{zhan04}. 
%The Mel cepstrum has proven to be one of the most successful feature 
%representations in speech related recognition tasks \cite{thom04}. 

In this paper 
we study the effect of resampling of speech  
on the MFCC parameters. 
We derive and show mathematically how the resampling of speech 
effects the extracted MFCC parameters and
establish a relationship between the MFCC parameters of 
resampled speech and that of the original speech. 
We focus our experiments primarily on the downsampled speech by a factor of 
$2$ and propose
six methods of computing the MFCC parameters of the downsampled speech, 
by an appropriate choice of the Mel-filter band, and compute the 
Pearson correlation between the MFCC of the original speech signal and the
computed MFCC of the down sampled speech to identify the best choice of the Mel
filter band.

%The motivation for this is as follows. 
%Compressed speech has many advantages including the reduction of storage space. 
%\comment{The speech recognition engine that has models of speech samples of a higher sampling rate (for e.g., 16kHz)
%can be tested with speech samples of lower sampling rate (for e.g., 8kHz) with the use of this
%relation. %If one can express the features of compressed speech in terms of the features of the original speech, 
%then one can store the features of compressed speech instead the uncompressed speech. 
%One can retrieve the features of uncompressed speech from the features of compressed speech at any given point of time. 
%This has a good application in Speech coding. 
%This reduces the storage space making the speech database of embedded devices more compact. 
%We also intend to make use of the compressed speech in speech recognition systems (instead of the uncompressed speech). 
%We expect that a speech recognition system can recognize the compressed speech up to a certain level 
%of compression and to find that limit of compression is a topic of interest.
%}
%The rest of the paper is organized as follows. 
%\review{
In Section \ref{sec:theory} 
we derive a relationship between the MFCC parameters computed for original
 speech
and the time scaled speech and discuss six different choice of Mel-filter bank
selection to the MFCC parameters of the downsampled speech.  
Section \ref{sec:experiments} 
gives the details of the experiments conducted to substantiate the derivation. 
We conclude in Section \ref{sec:conclusions}.
%}

%%%%%%%%%%%%%%%%%%%%%%%%%%%%%%%%%%%%%%%%%%%%%%%%%%%%%%%%%%%%%%%%%%%%%%%%%%%%%%%%%%%%%%%%%%%%
\section{Computing the MFCC parameters}
\label{sec:mfcc_theory}
%%%%%%%%%%%%%%%%%%%%%%%%%%%%%%%%%%%%%%%%%%%%%%%%%%%%%%%%%%%%%%%%%%%%%%%%%%%%%%%%%%%%%%%%%%%%
\begin{figure*}
\centering
\includegraphics[width=0.95\textwidth]{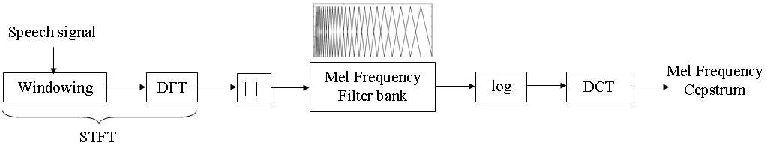}
\caption{Computation of Mel Frequency Cepstral Coefficients}
\label{fig:mfcc1}
\end{figure*}
The outline of the computation of Mel frequency cepstral coefficients (MFCC)
 is shown in Figure \ref{fig:mfcc1}. 
 %\comment{Can we make the figure better?} 
%Ignoring any spectral warping, the cepstral coefficients can be derived by Equation (\ref{eq:mfcc})
%\begin{equation}
%\label{eq:mfcc}
%c_{k} = \frac{1}{2\pi}\int_{-\pi}^{\pi}log|X(e^{j\omega})|.e^{j\omega k}d\omega
%\end{equation}
In general, the MFCCs are computed as follows. 
Let $x[n]$ be a speech signal with a sampling frequency of $f_s$, and is divided into $P$ 
frames each of length $N$ samples with an overlap of $N/2$ samples such that 
$\left \{ \vecx_{1}[n],  
\vecx_{2}[n] \cdots \vecx_{p}[n] \cdots \vecx_{P}[n] \right \} $, 
where $\vecx_{p}[n]$ denotes the $p^{th}$ frame of the speech signal
$x[n]$ and is
%\begin{equation}
$\vecx_p[n] = \left \{ x \left [p* \left (\frac{N}{2}-1 \right )+i \right ]  \right \}_{i=0}^{N-1}$
%\label{eq:signal_frame}
%\end{equation}
%Let the frame shift be $\frac{N}{2}$. 
Now the speech signal $x[n]$ can be represented 
in matrix notation as 
$ {\matrixx} \mydefine  
[\vecx_{1}, \vecx_{2}, \cdots, \vecx_{p}, \cdots, \vecx_{P}] $. %where
%\[
%\vecx_{p} = 
%\left [
%\begin{array} {c}
%x\left [(p-1)*\frac{N}{2}\right ] \\ 
%x\left [(p-1)*\frac{N}{2}+1 \right ]  \\ 
%\vdots \\
%x\left [(p-1)*\frac{N}{2}+N-1 \right ] 
%\end{array}
%\right ]
%\]
Note that the size of the matrix $\matrixx$ is $N \times P$. %\review{
The MFCC features
are computed for each frame of the speech sample (namely, for all $\vecx_{p}$).
%}
\subsection{Windowing, DFT and Magnitude Spectrum}
In speech signal processing, in order to compute the MFCCs of 
the $p^{th}$ frame, $\vecx_{p}$ is multiplied with a hamming window 
$w[n] = 0.54 - 0.46 \cos \left ( \frac{n \pi}{N} \right )$, followed by the discrete Fourier transform (DFT) as shown in (\ref{eq:dft}).
\begin{equation} 
X_{p}(k) = \sum_{n=0}^{N-1} x_p[n] w[n]\exp^{-j \frac{2\pi kn}{N}} 
\label{eq:dft} 
\end{equation} 
for $k=0, 1, \cdots, N-1$. 
If $f_{s}$ is the sampling rate of the speech signal $x[n]$ then
$k$ 
corresponds to the frequency $\linearf(k) = kf_{s}/N$.
Let $\vecX_{p} = [X_{p}(0), X_p(1),  \cdots,  X_{p}(N-1)]^T$ represent the 
DFT of the windowed $p^{th}$ frame of the speech signal $x[n]$, namely $\vecx_p$.
Accordingly, let $X = [\vecX_{1}, \vecX_{2}, \cdots \vecX_{p}, \cdots, \vecX_{P}]$ represent the DFT of the 
matrix $\matrixx$.
%\comment{
Note that the size of $X$ is $N \times P$ and is known as  STFT (short time
Fourier transform) matrix.
%}
%\subsection{Magnitude Spectrum}
The modulus of Fourier transform is extracted and the magnitude spectrum 
is obtained as $|X|$ which again is a matrix of size $N$ x $P$. 

\subsection{Mel Frequency Filter Bank}
The modulus of Fourier transform is extracted and the magnitude spectrum 
is obtained as $|X|$ which is a matrix of size $N \times P$. 
The magnitude spectrum is warped according to the Mel scale in order to 
adapt the frequency resolution to the properties of the human 
ear \cite{sirk01}. 
Note that the Mel ($\melf$) and the linear frequency ($\linearf$) \cite{thom04} are
related, namely, 
%\begin{equation}
%\label{eq:lf2mf}
$\melf = 2595 * log_{10}(1+\frac{\linearf}{700})$ 
%\end{equation}
where $\melf$ is the Mel frequency and $\linearf$ is the linear frequency. 
Then the magnitude spectrum $|X|$ 
is segmented into a number of critical bands by means of a 
Mel filter bank which typically consists of a series of overlapping 
triangular filters defined by their center frequencies 
$\linearf{_{c}}(m)$. 

The parameters that define a Mel filter bank are 
(a) number of Mel filters, $F$, (b) minimum frequency, $\linearf{_{min}}$ and 
(c) maximum frequency, $\linearf{_{max}}$. 
For speech, in general, it is suggested in \cite{link} 
that $\linearf{_{min}} > 100$ Hz. 
Furthermore, by setting $\linearf{_{min}}$ above 50/60Hz, we get rid of the 
hum resulting from the AC power, if present. 
\cite{link} also suggests that $\linearf{_{max}}$ be less than the 
Nyquist frequency. Furthermore, there is not much information above $6800$ Hz. 
Then a fixed frequency resolution in the Mel scale is computed using 
$\delta \melf = (\melf{_{max}}-\melf{_{min}})/(F+1)$ 
where $\melf{_{max}}$ and $\melf{_{min}}$ are the frequencies on the Mel scale corresponding 
to the linear frequencies $\linearf{_{max}}$ and $\linearf{_{min}}$ 
respectively. %which are computed using (\ref{eq:lf2mf}).
The center frequencies on the Mel scale are given by $\melf{_c(m)} = m.\delta
\phi$ where $m=1, 2, \cdots, F$.
To obtain the center frequencies of the triangular Mel filter bank in Hertz, we use the inverse relationship 
between $\linearf$ and $\melf$ given by $\linearf{_c(m)} =
700(10^{\melf{_c(m)}/2595}-1)$. The Mel filter bank, $M(m,k)$ \cite{pr_sigu06} is given by 
%\begin{eqnarray}
\[
\mbox{\scriptsize{$M(m,k)$}}=\left\{
\begin{array}{ll}
\mbox{\scriptsize{0}} & \mbox{\scriptsize{for $\linearf(k)<\linearf_{c}(m-1)$}} \\
\mbox{\scriptsize{$\frac{\linearf(k)-\linearf_c(m-1)}{\linearf_{c}(m)-\linearf_{c}(m-1)}$}} & \mbox{\scriptsize{for $\linearf_{c}(m-1)\le \linearf(k) < \linearf_{c}(m)$}} \\
\mbox{\scriptsize{$\frac{\linearf(k)-\linearf_c(m+1)}{\linearf_{c}(m)-\linearf_{c}(m+1)}$}} & \mbox{\scriptsize{for $\linearf_{c}(m)\le \linearf(k) < \linearf_{c}(m+1)$}} \\
\mbox{\scriptsize{0}} & \mbox{\scriptsize{for $ \linearf(k)\ge \linearf_{c}(m+1)$}}
\end{array} \right .
\]
%\end{eqnarray}
%where $f(k) = kf_s/N$. 
The Mel filter bank $M(m,k)$ is an $F \times N$ matrix. % which has a form shown below.

%\subsection{Log Mel Spectrum}
\subsection{Mel Frequency Cepstrum}
The logarithm of the filter bank outputs (Mel spectrum) is given in (\ref{ln}).
\begin{equation}
L_p(m,k) = ln\left \{\sum_{k=0}^{N-1}M(m,k) * |X_p(k)| \right \}
\label{ln}
\end{equation}
where $m = 1, 2, \cdots,  F$ and $p = 1, 2, \cdots, P$. 
The filter bank output, which is the product of the Mel filter bank, 
$M$ and the magnitude spectrum, $|X|$ is a $F \times P$ matrix. 
A discrete cosine transform of $L_p(m,k)$ results in the MFCC parameters. 
\begin{equation}
\label{dct}
\Phi_p^r \left \{ x[n] \right \} = \sum_{m=1}^{F}L_p(m,k) \cos \left \{\frac{r(2m-1)\pi}{2F} \right \}
\end{equation}
where $r = 1, 2, \cdots,  F$ and $\Phi_p^r \left \{ x[n] \right \}$ represents
the $r^{th}$ MFCC of the $p^{th}$ frame of the speech signal $x[n]$. 
The MFCC of all the $P$ frames of the speech signal are obtained as a matrix 
$\Phi$ 
\begin{equation}
\Phi \left \{\cal{X} \right \} = [\Phi_{1},  \Phi_{2}, \cdots,  \Phi_{p}, \cdots \Phi_{P}]
\end{equation}
Note that the $p^{th}$ column of the matrix $\Phi$, namely $\Phi_p$ represents the MFCC of the 
speech  signal, $x[n]$, corresponding to the $p^{th}$ frame, $x_p[n]$.
%%%%%%%%%%%%%%%%%%%%%%%%%%%%%%%%%%%%%%%%%%%%%%%%%%%%%%%%%%%%%%%%%%%%%%%%%%%%%%%%%%%%%%%%%%%%%%%%%%%%%%%%%%%%%%
\section{{MFCC of Resampled Speech}}
%Effect of speech resampling on MFCC}}
\label{sec:theory}
%%%%%%%%%%%%%%%%%%%%%%%%%%%%%%%%%%%%%%%%%%%%%%%%%%%%%%%%%%%%%%%%%%%%%%%%%%%%%%%%%%%%%%%%%%%%%%%%%%%%%%%%%%%%%%
In this section, we show how the resampling of the speech signal 
in time effects the computation of MFCC parameters. 
Let $y[s]$ denote the time scaled speech signal given by 
\begin{equation}
y[s] = x[\alpha n]  = x \downarrow \alpha
\end{equation} 
where $\alpha$ is the time scale modification (TSM) factor or the scaling factor
\footnote{We use $x[\alpha n]$ and 
$x \downarrow \alpha$ interchangeably. If $x = \left [1, 2, 3, ..., 2^n \right]_{1 X 2^n}$, 
then $x \downarrow 2 = [1, 3, 5, ... 2^n-1]_{1 X 2^{n-1}}$}. 
Let $y_{p}[s] = x_{p}[\alpha n]  = x_p \downarrow \alpha$ 
denote the $p^{th}$ frame of the time scaled speech where $s=0, 1,\cdots, S-1$, 
$S$ being the number of samples in the time scaled speech frame 
%The relation between $N$ and $S$ is 
given by $S=\frac{N}{\alpha}$. 
If $\alpha < 1$ the signal is expanded in time while
$\alpha > 1$ means the signal is compressed in time. 
Note that if $\alpha = 1$ the signal remains unchanged. 
%The DFT of the windowed $y_p[s]$ is given by
%\begin{equation} 
%Y_{p}(k') = \sum_{s=0}^{S-1} y_p[s] w[s]\exp^{-j \frac{2\pi k'n}{S}} \\
%\end{equation} 
%where $k' = 0, 1, \cdots, S-1$. 

DFT of the windowed $y_p[n]$ is calculated from the DFT of $x_p[n]$\footnote{For convenience, we ignore 
the effect of the window $w[n]$ on $y_p[n]$ or assume that $w[n]$ is also scaled by $\alpha$.}.
Assuming that $\alpha$ is an integer and using the scaling property of DFT \cite{open89}, %\comment{We have 
 we have,
\begin{equation}
Y_{p}(k') = \frac{1}{\alpha}\sum_{l=0}^{\alpha-1}X_{p}(k' + lS )
\label{eq:sampled_spectrum}
\end{equation}
where $k' = 1, 2, \cdots, S$.
The MFCC of the time scaled speech are given by
\begin{equation}
\label{eq:mfcctsm}
\Phi_p^r\{y[n]\} =  \Phi_{p}^{r} \left \{ x\downarrow\alpha \right \} 
= \sum_{m=1}^{F} L_p'(m,k') cos \left \{\frac{r(2m-1)\pi}{2F} \right \}
\end{equation}
%\begin{eqnarray}
%\Phi_p^r\{y[n]\} &=&  \Phi_{p}^{r} \left \{ x \downarrow \alpha \right \} \nonumber \\
% &=& \sum_{m=1}^{F} L_p'(m,k') cos \left \{\frac{r(2m-1)\pi}{2F} \right \}
%\label{eq:mfcctsm}
%\end{eqnarray}
where $r = 1, 2, \cdots, F$ and
\begin{equation}
L_p'(m,k') = ln \left \{\sum_{k'=0}^{S-1}\tiny{M'(m,k')}  \left |\frac{1}{\alpha}\sum_{l=0}^{\alpha-1}X_{p}(k' + lS)\right | \right \}
\label{eq:lns}
\end{equation}
Note that $L_p'$ and $M'$ are the log Mel spectrum and the Mel filter bank of the resampled speech. 
%We can rewrite $L_p'(m,k')$ in (\ref{eq:lns}) as 
%\begin{equation*}
%L_p'(m,k') = ln \left | \frac{1}{\alpha} \right | + ln \left \{\sum_{k'=0}^{S-1}M'(m,k') \left | \sum_{l=0}^{\alpha-1}X_{p}(k' + lS) \right | \right \} 
%\end{equation*}
%\begin{eqnarray*}
%L_p'(m,k') &=& ln \left \{\sum_{k'=0}^{S-1}\tiny{M(m,k')}  \left |\frac{1}{\alpha}\sum_{l=0}^{\alpha-1}X_{p}(k' + lS)\right | \right \} \\
%&=& ln \left | \frac{1}{\alpha} \right | + ln \left \{\sum_{k'=0}^{S-1}M(m,k') \left | \sum_{l=0}^{\alpha-1}X_{p}(k' + lS) \right | \right \} 
%\end{eqnarray*}
%The summation $\sum_{l=0}^{\alpha-1}X_{p}(k' + lS)$ can be expanded as
%\[
%\sum_{l=0}^{\alpha-1}X_{p}(k' + lS) = X_{p}(k')+X_p(k'+S)+ ... +X_p(k'+(\alpha-1)S)
%\]
%Using this, we get
%\begin{align}
%\label{eq:lp'2lp}
%L_p'(m,k') &= ln \left |\frac{1}{\alpha}\right | + L_p(m,k') + L_p(m,k'+S) + 
%\nonumber \\
%&\qquad ... + L_p(m,k'+(\alpha-1) S)
%\end{align}
%where $L_p(m,k')$ can be computed as ($\ref{ln}$). Substituting $L_p'(m,k')$ in (\ref{eq:mfcctsm}), we get the MFCC parameters of the time scaled speech. 
%Note that in the case when $\alpha = 1$, namely, there is no scaling of the speech signal. 
%Then $S=N$ and from (\ref{eq:lp'2lp}), we have, $L_p'(m,k') = L_p(m,k')$ where $k' = 0, 1, ..., N-1$. 
%Then (\ref{eq:lns}) reduces to (\ref{ln}) and $\Phi\{x\} = \Phi \{ x \downarrow 1 \}$.
We consider various forms of the Mel filter bank, $M'(m,k')$ which is used in the calculation of MFCC 
of the resampled speech. 
The best choice of the Mel filter band is the one 
which gives the best Pearson correlation between the 
MFCC of the original speech and the MFCC of the resampled speech. 

\subsection{Computation of MFCC of Resampled speech}
The major step in the computation of MFCC of the resampled speech 
lies in the construction of the 
Mel filter bank. 
The Mel filter bank used to calculate the MFCC of the resampled speech is given by  
\[
%\begin{tiny}
\mbox{\scriptsize{$M'(m,k')$}}=\left\{
\begin{array}{ll}
\mbox{\scriptsize{0}} & \mbox{\scriptsize{for $\linearf(k')<\linearf'_{c}(m-1)$}} \\
\mbox{\scriptsize{$\frac{\linearf(k')-\linearf'_c(m-1)}{\linearf'_{c}(m)-\linearf'_{c}(m-1)}$}}
& \mbox{\scriptsize{for $\linearf'_{c}(m-1)\le \linearf(k') < \linearf'_{c}(m)$}} \\
\mbox{\scriptsize{$\frac{\linearf(k')-\linearf'_c(m+1)}{\linearf'_{c}(m)-\linearf'_{c}(m+1)}$}}
& \mbox{\scriptsize{for $\linearf'_{c}(m)\le \linearf(k') < \linearf'_{c}(m+1)$}} \\
\mbox{\scriptsize{0}} & \mbox{\scriptsize{for $ \linearf(k')\ge \linearf'_{c}(m+1)$}}
\end{array} \right.
%\end{tiny}
\]
where $l_f(k') = \frac{k'(f_s/2)}{N/2}$.

As mentioned, we consider different 
forms of Mel filter banks and identify the 
Mel-filter bank
that results in the MFCC value of the resampled speech signal 
that matches best with the
original speech signal MFCC. This is done 
by computing the Pearson coefficient between
the MFCC of the resampled speech and the MFCC of the original speech.
The variations in the Mel filter banks 
is a result of the way in which the center frequencies 
and the amplitude of the filter coefficients are chosen. 
In all the cases discussed below, we assume, 
(a) $\alpha=2$, 
(b) the number of Mel filters used for the feature extraction of 
original speech and that of the resampled speech are same and, 
(c) the window length reduces by half, namely, $N/2$. 

\newcommand{\Aone}{A}
\newcommand{\Atwo}{B}
\newcommand{\C}{C}
\newcommand{\D}{D}
\newcommand{\Eone}{E}
\newcommand{\Etwo}{F}

\subsubsection{Type \Aone\ and Type \Atwo: Downsampling $M(m,k)$}
\label{sec:typea}

$M'(m,k')$ is obtained by downsampling $M(m,k)$ by a factor of $\alpha$, 
namely, $M_{A}'(m,k') = M(m, \alpha k)$. 
There are two ways in which the center frequencies of $M'_A$ are chosen.
{\em Type {\Aone}}: same as that of the original center frequencies, namely,
$\linearf'_{c}(m) = \linearf_{c}(m)$, and 
{\em Type {\Atwo}}: halving the original center frequencies, namely, 
$\linearf'_{c}(m) = \frac{1}{2}\linearf_{c}(m)$. 

%\subsubsection{Type B: Doubling the coefficients of downsampled $M(m,k)$}
%The filter coefficients obtained in {{\em Type A}} are doubled in magnitude, 
%namely, $M_{B}'(m,k') = 2*M(m, \alpha k)$. 
%\comment{The two cases discussed in {{\em Type A}} are considered here also. }

\subsubsection{Type \C: Constructing new filter bank in the halved band}

Here, we halve the frequency band on which the original filter bank ($M$) 
is constructed 
and construct a new filter bank following the steps described in 
Section (\ref{sec:mfcc_theory}) on the halved band. 
The minimum and maximum frequencies of the new Mel bank 
are chosen as $\frac{\linearf{_{max}}}{2}$ and $\frac{\linearf{_{min}}}{2}$
respectively.

\subsubsection{{Type \D: Interpolating}}
Here, alternate center frequencies of the original Mel bank are halved 
and filters are constructed with the resultant center frequencies. 
This reduces the bandwidth of the Mel bank and the number of Mel filters by a
factor of $2$. 
The output of these $\frac{F}{2}$ Mel filters are denoted as 
$[\vec{g_1} \vec{g_2} \cdots \vec{g_m} \cdots  \vec{g_{F/2}}]$. 
and the Mel spectrum is computed as 
\[
\left [ \vec{g_1} \;\; \frac{\vec{g_1}+\vec{g_2}}{2}\;\;  \vec{g_2}  \;\;
\frac{\vec{g_2}+\vec{g_3}}{2} \cdots \vec{g_m}  \cdots  \vec{g_{F/2}} \;\; 
\frac{\vec{g_{F/2}}+\vec{g_1}}{2} \right ]
\]
DCT of the logarithm of the above vectors gives the MFCC of the down sampled speech. 

\subsubsection{{Type \Eone\ and Type \Etwo: Reversing, Adding and Averaging }}
\label{sec:typee}
\begin{figure}[t]
\centerline{\epsfig{figure=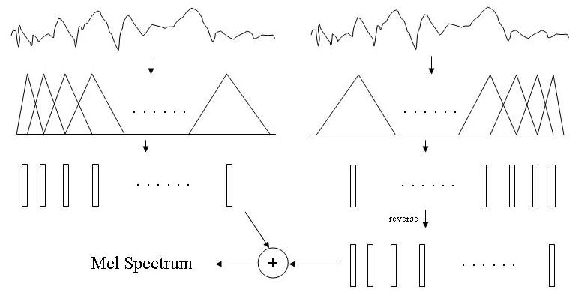,width=0.85\textwidth}}
\caption{{\it Type \Eone\ and \Etwo\ - Reversing, Adding and Averaging.}}
\label{fig:E}
\end{figure}

In this case, the filter bank outputs of the downsampled Mel filter bank, 
namely, $M_{A}'(m,k')$ are computed. 
Then the downsampled Mel filter bank is mirrored/reversed such that the filter with the 
highest bandwidth 
comes first and the one with the lowest bandwidth comes last. 
The spectrum of the downsampled signal is passed through this reversed filter bank 
and the filter bank outputs are again reversed. %such that the output of filter with lowest bandwidth . 
These reversed filter bank outputs are added to the former filter bank (downsampled bank) outputs 
and their average is considered to be the Mel spectrum. 
DCT of the logarithm of the Mel spectrum gives the MFCC of the down sampled speech. 
This method also has $2$ cases, namely, 
{\em Type \Eone}: the center frequencies chosen are of type {\em Type \Aone}, and, 
{\em Type \Etwo}: the center frequencies chosen are of type {\em Type \Atwo}.
%This process is depicted in Figure (\ref{fig:E}).
This process is depicted in Figure \ref{fig:E}.

%%%%%%%%%%%%%%%%%%%%%%%%%%%%%%%%%%%%%%%%%%%%%%%%%%%%%%%%%%%%%%%%%%%%%%%%%%%%%%%%%%%%%%%%%%%%
\section{Experimental Results}
\label{sec:experiments}
%%%%%%%%%%%%%%%%%%%%%%%%%%%%%%%%%%%%%%%%%%%%%%%%%%%%%%%%%%%%%%%%%%%%%%%%%%%%%%%%%%%%%%%%%%%%
In all our experiments we considered speech signals sampled at $16$ kHz and
represented by $16$ bits. The speech signal is divided into frames of duration $32$ ms 
(or $N = 512$ samples) and $16$ ms overlap ($256$ samples). 
MFCC parameters are computed for each speech frame using (\ref{dct}). 
%Initially the speech frame is pre processed by passing it through a hamming window and 
%the magnitude spectrum of
%the Fourier transform is warped according to the Mel scale.
The Mel filter bank used has $F=30$ bands spread from $\linearf{_{min}} = 130$ Hz 
%(minimum frequency) 
to a maximum frequency of  $\linearf{_{max}} = 6800$ Hz. 
The MFCC parameters 
(denoted by $\Phi \{x[n]\} = %\Phi \left \{\cal{X} \right \} = 
[\Phi_{1},  \Phi_{2}, \cdots,  \Phi_{m}, \cdots \Phi_{F}]$\footnote{Note that $\Phi_m$ is a vector formed with the $m^{th}$ MFCC of all the speech frames}) are computed for 
the $16$ kHz speech signal $x[n]$, as described in %\comment{Check Section Number} 
Section \ref{sec:mfcc_theory}. %(see(\ref{dct})). 
Then $x[n]$ is downsampled by a scaling factor of $\alpha=2$ and 
denoted by $y[s] = x \downarrow 2 = x[2n]$. 
The MFCC parameters of $y[s]$ (denoted by $\Phi \{y[s]\} = %\Phi \left \{\cal{X} \right \} = 
[\Phi'_{1},  \Phi'_{2}, \cdots,  \Phi'_{m}, \cdots \Phi'_{F}]$) 
are calculated using the six methods discussed in Sections
\ref{sec:typea} to \ref{sec:typee}. 
Pearson correlation coefficient (denoted by $r$)
\footnote{Pearson correlation coefficient between two 
vectors $\vec{X}$ and $\vec{Y}$ %\cite{ } 
each of length $n$ is given by
\[
r = \frac{\sum{\vec{X}\vec{Y}}-\frac{1}{n}\sum{\vec{X}}\sum{\vec{Y}}} {\sqrt{(\sum{\vec{X}^2}-\frac{1}{n}(\sum{\vec{X}})^2) (\sum{\vec{Y}^2}-\frac{1}{n}(\sum{\vec{Y}})^2)}}
\].} is computed between the MFCC parameters of the downsampled speech (using
different Mel-filter bank constructs) and the MFCC of the original speech  
in two different ways.

\begin{figure}[t]
\centerline{\epsfig{figure=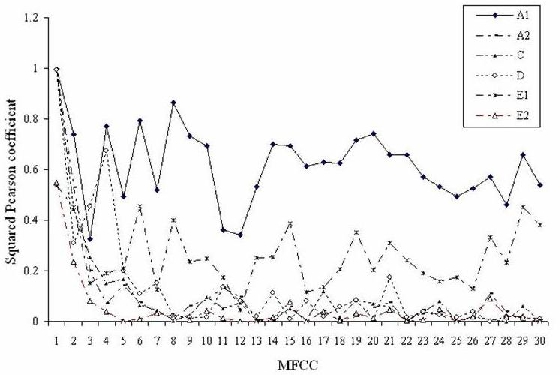,width=0.85\textwidth}}
\caption{{\it Pearson correlation ($r^2$) between the MFCC of original speech and 
downsampled speech (for speech sample 3).}}  
\label{fig:r}
\end{figure}

\noindent {{\bf Case I}}: 
Pearson correlation coefficient, $r$ between the individual MFCC of the original and 
the downsampled speech signals,
namely, $\Phi_{m}$ and $\Phi'_{m}$, $m=1, 2, \cdots, F$ is calculated. 
The variation of the squared Pearson correlation coefficient, $r^2$ over
individual MFCC ($F=30$) for the $6$ types of Mel filter bank constructs 
is shown in 
Figure \ref{fig:r}. 

\noindent {{\bf Case II}}: 
The $F$ MFCC vectors are concatenated to form a single vector and the $r$ between the 
two vectors corresponding to 
the original speech and the downsampled speech is computed. 
The Pearson correlation coefficient, $r$ for the $6$ methods 
is shown in Table \ref{tab:r} for three different $16$ kHz, $16$ bit speech samples. 
\begin{table} [t,h]
\caption{\label{tab:r} {\it Pearson correlation ($r$) 
between the MFCC of original speech and the downsampled
speech}}
\centerline{
\begin{tabular}{|c|cccccc|}
\hline
%Speech & & & Type & & &\\
Speech & \Aone & \Atwo & \C & \D & \Eone & \Etwo\\
\hline  
%1 & 0.957 & 0.899 & 0.888 & 0.813 & 0.871 & 0.839\\
Sample 1 & {\bf 0.978} & 0.945 & 0.941 & 0.908 & 0.844 & 0.821\\
Sample 2 & {\bf 0.976} & 0.947 & 0.943 & 0.914 & 0.889 & 0.877\\
Sample 3 & {\bf 0.973} & 0.947 & 0.944 & 0.916 & 0.895 & 0.878\\
\hline
\end{tabular}}
\end{table}

As observed from Figure \ref{fig:r} and Table \ref{tab:r}, 
the {\em Type \Aone} of constructing Mel filter bank for the down sampled speech 
gives the best correlation between the MFCC parameters of the original speech and 
that of the downsampled speech.

%%%%%%%%%%%%%%%%%%%%%%%%%%%%%%%%%%%%%%%%%%%%%%%%%%%%%%%%%%%%%%%%%%%%%%%%%%%%%%%%%%%%%%%%%%%%
\section{Conclusion}
\label{sec:conclusions}
The effect of resampling of speech on the MFCC parameters of speech has been
presented. 
We have demonstrated that it is possible to extract MFCC from a downsampled
speech by constructing an appropriate  Mel filter bank. We presented 
six methods of computing MFCC of a downsampled speech signal by transforming 
the Mel filter bands used to compute MFCC parameters. The choice of
various transformation of Mel filter bank was based on the relationship between
the spectrum of the original and the resampled signal (Equation \ref{eq:sampled_spectrum}).
We have shown that the Pearson correlation coefficient between the MFCC parameters of the 
original speech and the downsampled speech shows a good fit with a downsampled
version of the Mel filter bank ({\em Type \Aone}). We believe the results
presented in this paper will enable us to experiment and measure the performance of a 
speech recognition engine (statistical phoneme models derived from original speech) 
on subsampled speech (time compressed speech).

%%%%%%%%%%%%%%%%%%%%%%%%%%%%%%%%%%%%%%%%%%%%%%%%%%%%%%%%%%%%%%%%%%%%%%%%%%%%%%%%%%%%%%%%%%%%

%%%%%%%%%%%%%%%%%%%%%%%%%%%%%%%%%%%%%%%%%%%%%%%%%%%%%%%%%%%%%%%%%%%%%%%%%%%%%%%%%%%%%%%%%%%%
%\section{Acknowledgements}
%%%%%%%%%%%%%%%%%%%%%%%%%%%%%%%%%%%%%%%%%%%%%%%%%%%%%%%%%%%%%%%%%%%%%%%%%%%%%%%%%%%%%%%%%%%%

%\eightpt
\bibliographystyle{IEEEtran}
\bibliography{TSM_MFCC}

\end{document}